\documentclass{article}

\usepackage{arxiv}
\usepackage{color}
\usepackage[utf8]{inputenc} 
\usepackage[T1]{fontenc}    
\usepackage{hyperref}       
\usepackage{url}            
\usepackage{booktabs}       
\usepackage{amsfonts}       
\usepackage{nicefrac}       
\usepackage{microtype}      
\usepackage{lipsum}		
\usepackage{graphicx}
\usepackage{natbib}
\usepackage{doi}
\usepackage{amsmath} 
\usepackage{amssymb}
\usepackage{multirow}

\title{Stability, electronic disruption, and anisotropic superconductivity of hydrogenated trilayer metal tetraborides (MB$_{4}$H; M=Be, Mg, Ca, Al)}

\author{ \href{https://orcid.org/0009-0004-2196-8245}{\includegraphics[scale=0.06]{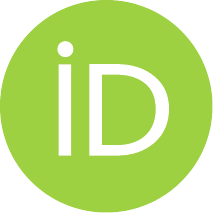}\hspace{1mm}Jakkapat Seeyangnok} \\
	Department of Physics\\
    Faculty of Science\\
	Chulalongkorn University\\
	Bangkok, Thailand \\
	\texttt{jakkapatjtp@gmail.com} \\
	\And
	\href{https://orcid.org/0000-0002-8450-7751}{\includegraphics[scale=0.06]{orcid.pdf}\hspace{1mm}Udomsilp Pinsook*} \\
	Department of Physics\\
    Faculty of Science\\
	Chulalongkorn University\\
	Bangkok, Thailand \\
	\texttt{Udomsilp.P@Chula.ac.th} \\
    \And
	\href{https://orcid.org/0000-0002-1205-7675}{\includegraphics[scale=0.06]{orcid.pdf}\hspace{1mm}Graeme J. Ackland} \\
	Centre for Science at Extreme Conditions,\\
    School of Physics and Astronomy,\\
	University of Edinburgh,\\
	Edinburgh, United Kingdom. \\
	\texttt{gjackland@ed.ac.uk} \\
}

\begin{document}
\maketitle

\begin{abstract}
The discovery of superconductivity in MgB$_2$ (\(T_c = 39\) K) \cite{nagamatsu2001superconductivity} established metal diborides (MB$_2$) as a promising class of conventional superconductors. Recent advances in fabrication techniques have enabled the synthesis of 2D MgB$_2$ with a \(T_c\) of 36 K \cite{cheng2018fabrication}, reigniting interest in layered metal borides. This has led to predictions of superconductivity in various 2D metal borides, including MB$_4$ (M = Be, Mg, Ca, Al), with CaB$_4$ exhibiting the highest estimated \(T_c\) of 36.1 K. To explore the impact of hydrogenation on superconductivity, we systematically investigate two-dimensional hydrogenated trilayer metal borides (MB$_4$H; M = Be, Mg, Ca, Al). Our results reveal that these materials retain a metallic nature dominated by boron \(p\)-orbitals, while hydrogenation significantly alters their band dispersion and Fermi surface topology. Phonon calculations confirm their dynamical stability and reveal strong electron-phonon interactions, leading to multi-gap superconductivity. Among the studied compounds, MgB$_4$H, AlB$_4$H, and CaB$_4$H exhibit possible two superconducting gaps, with CaB$_4$H showing the strongest electron-phonon coupling, resulting in an intrinsic superconducting transition temperature of 64 K. In contrast, AlB$_4$H shows the weakest coupling, with \(T_c = 22\) K. The calculated electron-phonon coupling constants (\(\lambda\)) range from 0.62 to 0.99, demonstrating the tunability of superconducting properties through elemental substitution. These findings provide valuable insights into superconductivity in hydrogenated metal borides and highlight their potential for high-\(T_c\) applications.
\end{abstract}


	\section{Introduction} \label{Sec:Intro}
The search for high-temperature superconductors has been a central theme in condensed matter physics, particularly in hydrogen-rich materials, ever since metallic hydrogen was proposed as a potential high-$T_c$ superconductor \cite{ashcroft2004hydrogen, mcmahon2011high}. Breakthroughs in this field include the discovery of record-high superconducting transition temperatures in hydrogen-based compounds, such as H$_3$S ($T_c \sim 200$ K) and LaH$_{10}$ ($T_c \sim 280$ K) \cite{duan2014pressure, peng2017hydrogen, liu2017potential}. However, these materials require extreme pressures in the range of 150–180 GPa, limiting their practical applications \cite{drozdov2015conventional, einaga2016crystal, drozdov2019superconductivity, somayazulu2019evidence}. As a result, attention has shifted toward hydrogenated materials that remain stable at ambient pressure. 

Boron has long been recognised as a promising element for superconductivity: like hydrogen it benefits from a low mass, but is non-metallic under ambient conditions.  Pressure can break the covalent bonds, 
becoming metallic  and superconducting at 160 GPa, with T$_c$ around 6K at 175 GPa, rising to 11.2K at 250 GPa \cite{Eremets2001Superconductivity}.  Metallicity and superconductivity due to boron can be maintained to ambient conditions in the compound MgB$_2$, in which electrons donated from the cation prevent the formation of closed-shell covalent bonding in the 2D honeycomb Boron layers.  Combining the high temperatures of hydride superconductors with the ambient stability of boron-based superconductors is a promising path toward  superconductivity at ambient conditions.  It is interesting to note that the recently-proposed room-temperature superconductor LaSc$_2$H$_{24}$ has the MgB$_2$ structure with added hydrogen, and trivalent Sc in place of trivalent B.\cite{He2024LaSc2H24,Song2025LaSc2H24Experiment}. Recent studies have demonstrated that hole-doped Mg(BH$_4$)$_2$ can become metallic and exhibit superconductivity with a $T_c$ of 140 K \cite{liu2024realizing}, while Mg$_2$IrH$_6$ has been reported to reach $T_c$ = 160 K \cite{dolui2024feasible}, further motivating research into low-dimensional hydrogenated superconductors.

Among the various superconducting systems, metal borides have garnered significant interest due to their strong electron-phonon coupling and high potential $T_c$. The discovery of superconductivity in MgB$_2$ ($T_c = 39$ K) \cite{nagamatsu2001superconductivity} established metal diborides (MB$_2$) as promising conventional superconductors along with recent advancements in fabrication techniques have enabled the synthesis of 2D MgB$_2$ with a $T_c$ of 36 K \cite{cheng2018fabrication}, fueling renewed interest in layered metal borides. Since then, theoretical investigations have predicted superconductivity in several 2D metal borides, including MB$_4$ (M = Be, Mg, Ca, Sc, Al), with estimated $T_c$ values reaching 36.1 K for CaB$_4$ while AlB$_4$, BeB$_4$ and MgB$_4$ have T$_c$ of 30.9 K, 29.9 K, and 22.2 K, respectively \cite{sevik2022high}. Several boride compounds can be considered as a superhard material, due to high hardness value and high phonon frequencies \cite{superhard1,superhard2}. However, these compounds are often found to be an insulator, because of strong bonding between B-B atoms. 

The incorporation of hydrogen into two-dimensional (2D) materials has emerged as an effective strategy for enhancing superconducting properties, as hydrogen-derived bands can metallize boride and related compounds. Hydrogenated materials can generally be classified into two major types. (i) Substitutional hydrogenation, where hydrogen atoms replace specific elements—such as in Janus transition-metal dichalcogenide hydrides (TMDHs)—has been motivated by the successful synthesis of the MoSH monolayer via the SEAR technique~\cite{lu2017janus} and the predicted superconductivity with $T_c \approx 27$~K~\cite{liu2022two} accompanied by a possible charge-density-wave (CDW) instability~\cite{ku2023ab}. Subsequently, several hydrogenated tungsten analogues have been theoretically proposed, exhibiting $T_c$ values around 12~K~\cite{seeyangnok2024superconductivity,seeyangnok2024superconductivitywseh,qiao2024prediction}, with CDW features similar to those in MoSH and consistent with later reports on their phase stability and superconductivity~\cite{gan2024hydrogenation,fu2024superconductivity}. Further studies have extended to group-IV transition-metal dichalcogenides (M = Ti, Zr, Hf; X = S, Se, Te), predicting superconducting transitions in the 9–30~K range~\cite{li2024machine,ul2024superconductivity}. However, many Janus MXH monolayers exhibit magnetic ground states~\cite{seeyangnok2025competition}, reminiscent of CrSH~\cite{sukserm2025half}, highlighting the complex interplay between magnetism and superconductivity in substitutionally hydrogenated 2D materials.

(ii) Direct hydrogenation preserves the parent lattice while integrating hydrogen atoms into interstitial or surface sites. Representative systems include hydrogenated graphene ($T_c > 90$~K)~\cite{savini2010first}, Mo$_2$C$_3$ ($T_c = 53$~K)~\cite{jiao2022hydrogenation}, and CuH$_2$ ($T_c > 40$~K)~\cite{yan2022enhanced}. Theoretical predictions further indicate that hydrogenated MgB$_2$ could reach $T_c = 67$~K, increasing up to 100~K under biaxial strain~\cite{bekaert2019hydrogen}, while hydrogenated phosphorus carbide (HPC$_3$) may achieve $T_c = 57.3$~K under tensile strain~\cite{li2022phonon}. Similarly, 2D hydrogenated metal diborides (M$_2$B$_2$H; M = Al, Mg, Mo, W) have been predicted to exhibit superconductivity in the range 18–52~K~\cite{han2023theoretical}, with Ti$_2$B$_2$H$_4$ reaching $T_c = 69.4$~K under tensile strain~\cite{han2023high} and potentially up to 84~K according to recent investigations~\cite{seeyangnok2025high_npj}.

In this work, we conduct a systematic investigation into the superconducting properties of two-dimensional hydrogenated trilayer metal borides (MB$_4$H; M = Be, Mg, Ca, Al). Our study explores their crystal structure, electronic properties, and superconducting state mediated by phonon-driven electron-electron interactions. We reveal a strong interplay between electronic anisotropy, multi-gap superconductivity, and phonon-mediated coupling, shedding light on the fundamental mechanisms governing superconductivity in 2D hydrogenated metal borides. These findings contribute to the broader understanding of hydrogen-enhanced superconductivity in low-dimensional systems and offer insights into the design of next-generation superconducting materials.

\section{Methodology} \label{Sec:Theory}
The structural and electronic properties of MB$_4$H (M = Be, Mg, Ca, Al) were investigated using density functional theory (DFT) as implemented in \textsc{Quantum Espresso} (QE) \cite{giannozzi2009quantum, giannozzi2017advanced}. The crystal structures were modeled using \textsc{VESTA} \cite{momma2011vesta} and fully optimized via the Broyden–Fletcher–Goldfarb–Shanno (BFGS) algorithm \cite{BFGS, liu1989limited}. Structural relaxation was performed until the force on each atom converged below $10^{-5}$ eV/\AA, with a vacuum thickness of $20$ \AA \ to prevent spurious interactions between periodic images.  For exchange-correlation effects, we employed the generalized gradient approximation (GGA) with the Perdew–Burke–Ernzerhof (PBE) functional \cite{perdew1996generalized} along with optimized norm-conserving Vanderbilt (ONCV) pseudopotentials \cite{hamann2013optimized, schlipf2015optimization}. The plane-wave basis set was truncated at 80 Ry for the wavefunction and 320 Ry for the charge density. Brillouin zone sampling for structural relaxation was performed using a $24\times24\times1$ Monkhorst-Pack k-point grid \cite{monkhorst1976special}, incorporating a Marzari-Vanderbilt-DeVita-Payne cold smearing of $0.02$ Ry \cite{marzari1999thermal} to improve convergence. The Fermi surface was visualized with \textsc{XCRYSDEN} \cite{kokalj2003computer}.  

Dynamical stability was assessed through phonon calculations using density functional perturbation theory (DFPT) with a $24\times24\times1$ k-point grid and a $12\times12\times1$ q-point grid. Electron-phonon interactions were computed using the Wannier-Fourier interpolation method \cite{giustino2017electron, giustino2007electron}, as implemented in the \textsc{EPW} package \cite{noffsinger2010epw, ponce2016epw}. This method enables precise determination of the electron-phonon coupling constant ($\lambda$) and the critical superconducting temperature ($T_c$).  To investigate superconducting behavior, we employed the anisotropic Migdal-Eliashberg (ME) theory \cite{margine2013anisotropic} by solving the coupled nonlinear equations:  

\begin{equation}  
Z_{nk}(i\omega_{j}) = 1+\frac{\pi T}{N({\varepsilon_{F})\omega_{j}}}\sum_{mk'j'}\frac{\omega_{j'}}{\sqrt{\omega^{2}_{j'}+\Delta^{2}_{mk'}(i\omega_{j'})}},  
\end{equation}  

\begin{align}  
Z_{nk}(i\omega_{j})\Delta_{mk}(i\omega_{j}) &= \frac{\pi T}{N(\varepsilon_{F})} \sum_{mk'j'} \frac{\Delta_{mk'}(i\omega_{j'})}{\sqrt{\omega^{2}_{j'}+\Delta^{2}_{mk'}(i\omega_{j'})}} \\  
& \times[\lambda(nk,mk',\omega_j -\omega_{j'})-\mu^*]\delta(\epsilon_{mk'}-\varepsilon_F).  
\end{align}  

These equations were solved self-consistently along the imaginary axis at fermionic Matsubara frequencies $\omega_j = (2j+1)\pi T$. To ensure numerical accuracy, we used $120\times120\times1$ k-point and $60\times60\times1$ q-point grids in the Wannier-Fourier interpolation. The convergence of $\lambda$ was confirmed by analyzing the stability of $\alpha^2F(\omega)$ and $\lambda(\omega)$ with increasing k- and q-point densities.  Additional computational parameters included a Fermi surface thickness of $0.64$ eV, a Matsubara frequency cutoff of $1.6$ eV, and Gaussian broadening with widths of $0.16$ eV for electrons and $0.5$ meV for phonons. The Morel-Anderson pseudopotential ($\mu^*$) was set to 0.1 to account for Coulomb repulsion.  

To evaluate the mechanical stability of two-dimensional MB\(_{4}\)H materials, we calculated the in-plane elastic constants based on strain–energy relationships appropriate for 2D hexagonal systems. The elastic constants \(C_{ij}\) were obtained from the second derivatives of the total energy with respect to applied strain components as follows:
\begin{equation}
C_{ij} = \frac{1}{S_0} \frac{\partial^2 E}{\partial \epsilon_i \partial \epsilon_j},
\end{equation}
where \(S_0\) is the equilibrium area of the unit cell, and \(\epsilon_i\), \(\epsilon_j\) denote the in-plane strain components.

In hexagonal lattices, the in-plane elastic constants reduce to two independent parameters: \(C_{11}\) and \(C_{12}\). The shear modulus is defined as \(C_{66} = \frac{C_{11} - C_{12}}{2}\). These constants describe the linear stress--strain relationship, which in matrix form is expressed via the generalized Hooke’s law for 2D materials:
\begin{equation}
\sigma =
\begin{bmatrix}
C_{11} & C_{12} & 0 \\
C_{12} & C_{22} & 0 \\
0 & 0 & C_{66}
\end{bmatrix}
\varepsilon.
\end{equation}

The elastic constants were computed by applying small, in-plane deformations to the relaxed structure, followed by fitting the resulting energy–strain curves to extract second-order coefficients.

\section{Phase stability}
\subsection{Crystal structure}
        \begin{figure}[tbh!]
        \centering
		\includegraphics[width=9cm]{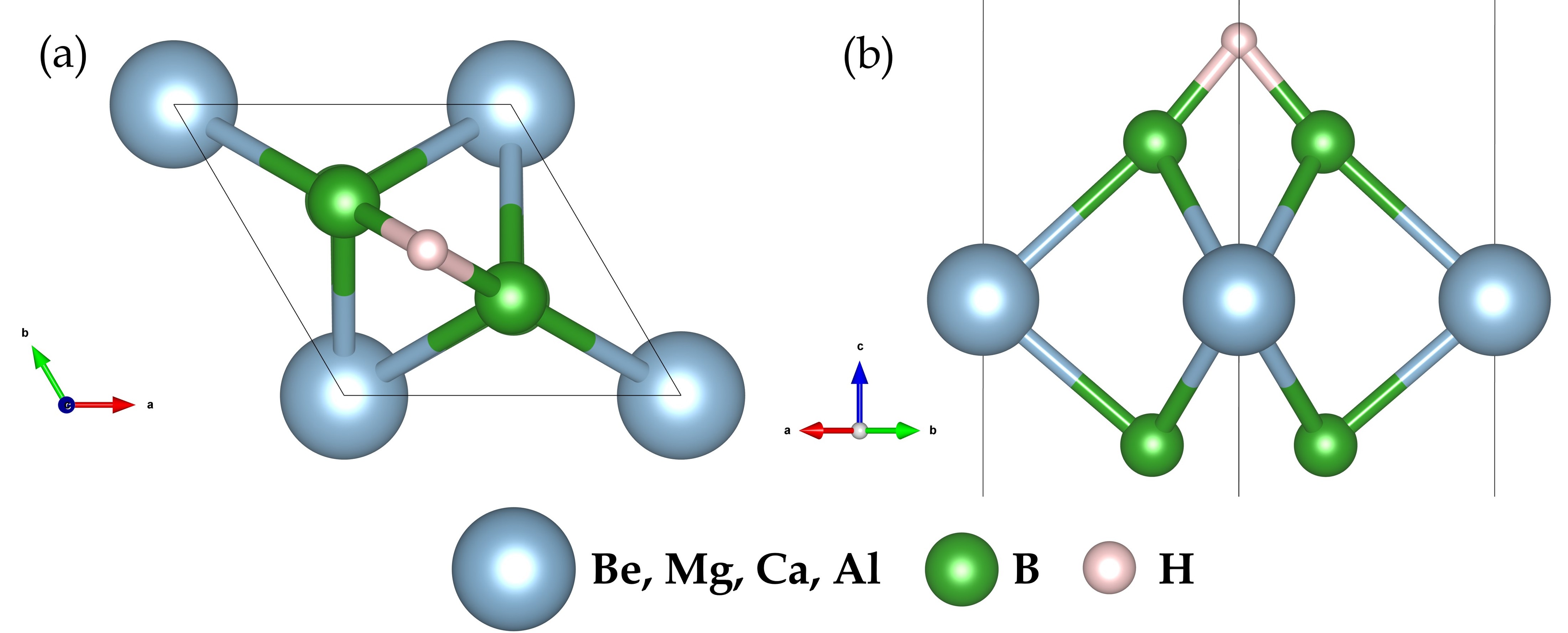}
		\caption{Figures (a) and (b) show side and top views of the 2D MB$_{4}$H structure; M=Be, Mg, Ca, Al where metals (M), boron (B) and hydrogen (H) atoms are represented by blue, green and pink spheres, respectively.}
		\label{fig:2D-m2b2h4}
    \end{figure}
The crystal structures of the 2D MB$_{4}$H compounds (M = Be, Mg, Ca, Al) are obtained by hydrogenating a hydrogen atom onto the top layer of a trilayer metal diboride, which adopts a 3D hexagonal structure with space group symmetry $P6/mmm$. In this structure, metal atoms form a honeycomb lattice at the Wyckoff positions (0,0), while boron atoms occupy the positions (1/3,2/3) and (2/3,1/2). The hydrogenation position follows a similar trend to previous investigations of hydrogenated boride monolayers \cite{meninno2022absence} at (1/2, 1/2).  

\begin{table}[tbh!]
    \centering
    \caption{Structural parameters of 2D MB$_4$H monolayers, including the lattice constant ($a$), the thickness between the two boron layers ($h_1$), the thickness between the top boron layer and the hydrogen atom ($h_2$). The bond lengths are defined as follows: M--B(up) and M--B(down) denote the distances between the metal atom and the boron atoms in the upper and lower boron layers, respectively; 
    B--B(up/down) represents the B--B bond lengths within the upper and lower boron layers; and B--H is the bond length between the topmost boron and the hydrogen atom, as illustrated in Fig.~\ref{fig:2D-m2b2h4}.}
    \begin{tabular}{|c|c|c|c|c|c|c|c|}
        \hline
        2D compounds & $a$ (\AA) & $h_1$ (\AA) & $h_2$ (\AA) & M--B(up) (\AA) & M--B(down) (\AA) & B--B(up/down) (\AA) & B--H (\AA) \\
        \hline
        BeB$_{4}$H & 2.96 & 2.67 & 1.06 & 2.24 & 2.15 & 1.66 / 1.73 & 1.34 \\
        MgB$_{4}$H & 3.03 & 3.52 & 1.03 & 2.53 & 2.47 & 1.68 / 1.76 & 1.33 \\
        CaB$_{4}$H & 3.09 & 4.07 & 1.01 & 2.74 & 2.68 & 1.72 / 1.79 & 1.33 \\
        AlB$_{4}$H & 3.01 & 3.09 & 1.03 & 2.38 & 2.27 & 1.71 / 1.77 & 1.34 \\
        \hline
    \end{tabular}
    \label{tab:lattice-consts}
\end{table}

The fully optimized structures reveal that the top layer of boron slightly shifts toward the hydrogen atom. The lattice parameters ($a$) for different metals are 2.96, 3.03, 3.09, and 3.01~\AA\ for Be, Mg, Ca, and Al, respectively. The thickness ($h_1$) between the two boron layers is 2.67, 3.52, 4.07, and 3.09~\AA\ for Be, Mg, Ca, and Al, respectively. The thickness ($h_2$) between the top boron layer and the hydrogen atom is 1.06, 1.03, 1.01, and 1.03~\AA\ for Be, Mg, Ca, and Al, respectively. All of this information is summarized in Table~\ref{tab:lattice-consts}.

    \subsection{Energetic stability}



To assess the feasibility of synthesizing hydrogenation of MB$_4$ (M = Be, Mg, Ca, Al), we evaluated their energetic stability by calculating the formation energy with respect to the pristine MB$_4$ phase, given by

\begin{equation}
    E_{f} = E(\mathrm{MB_4H}) - E(\mathrm{MB_4}) - E(\mathrm{H_2})/2,
\end{equation}

where \(E(\mathrm{MB_4H})\), \(E(\mathrm{MB_4})\), and \(E(\mathrm{H_2})\) are the total energies of the hydrogenated monolayer, the pristine MB$_4$ monolayer, and a hydrogen molecule, respectively. The negative $E_f$ values listed in Table~\ref{tab:formation-energy} indicate that hydrogenation of MB$_4$ is energetically favorable, suggesting that if the pristine MB$_4$ monolayers can be synthesized, the formation of MB$_4$H should be thermodynamically feasible.

\begin{table}[tbh!]
    \centering
    \caption{Formation energies ($E_f$) of MB$_4$H (M = Be, Mg, Ca, Al) monolayers relative to their pristine MB$_4$ and H$_2$ molecule. Negative values indicate that hydrogenation is energetically favorable.}
    \begin{tabular}{|c|c|}
        \hline
        2D compounds & Formation energies (eV/formula unit) \\
        \hline
        BeB$_4$H & $-0.1781$ \\
        MgB$_4$H & $-0.5457$ \\
        CaB$_4$H & $-0.8939$ \\
        AlB$_4$H & $-0.3952$ \\
        \hline
    \end{tabular}
    \label{tab:formation-energy}
\end{table}

    \subsection{Thermal stability}

   \begin{figure}[h!]
		\centering
		\includegraphics[width=16cm]{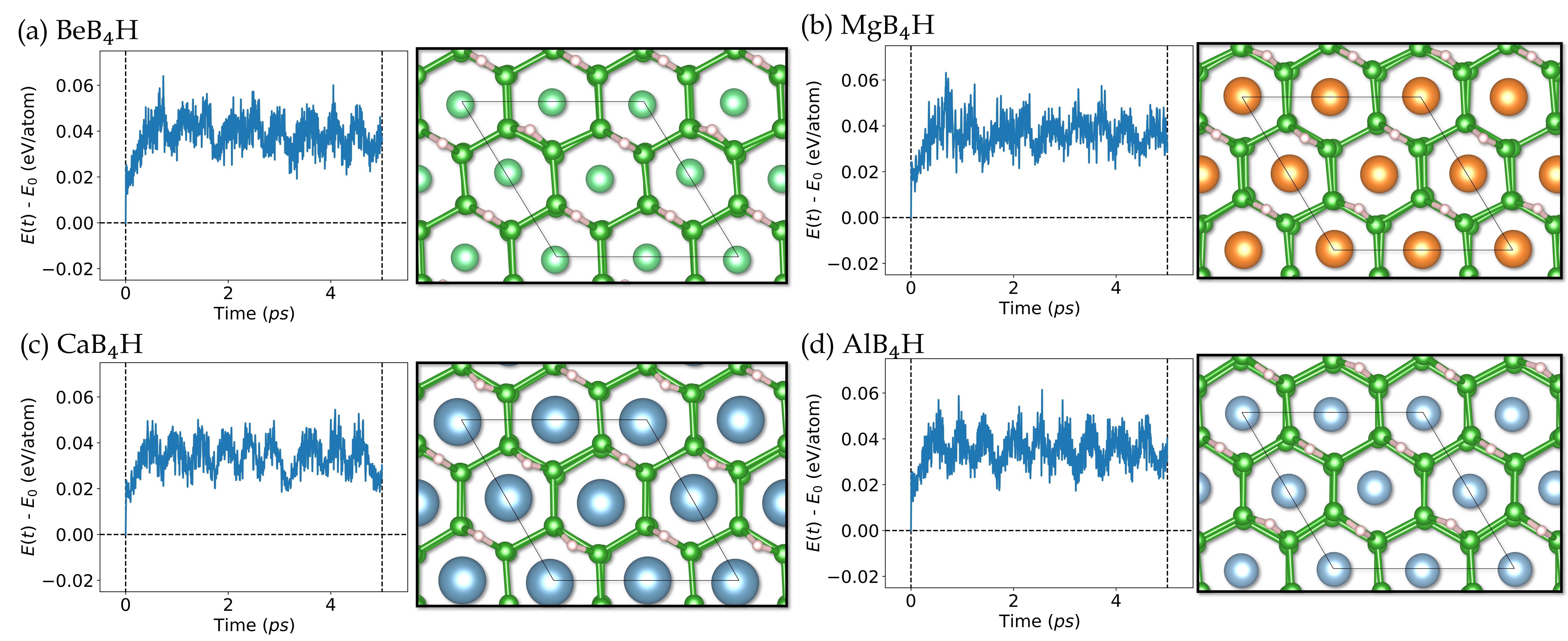}
		\caption{(a–d) Time evolution of the total energy fluctuation per atom, $E(t) - E_0$, during a 5 ps *ab initio* Born–Oppenheimer molecular dynamics (BOMD) simulation in the NVT ensemble at 300 K for a $2 \times 2 \times 1$ supercell (24 atoms) of the 2D MB$_{4}$H structure, where M = Be, Mg, Ca, and Al. The insets show top views of the final atomic configurations after 5 ps, demonstrating the dynamic structural stability of the systems.}
  	\label{fig:MD-plos}
	\end{figure}
    In addition to dynamical stability of phonon, which will be discussed further, we evaluated the thermal stability of the 2D MB$_{4}$H compounds (M = Be, Mg, Ca, Al), we employed \textit{ab initio} molecular dynamics (AIMD) simulations at finite temperature. Unlike phonon dispersion or elastic constant analysis—which are limited to the harmonic regime—AIMD offers a direct probe of the system’s dynamical behavior under thermal agitation, capturing potential structural rearrangements and anharmonic effects that may arise at room temperature.
    
    Simulations were carried out in the canonical (NVT) ensemble using a $2 \times 2 \times 1$ supercell containing 24 atoms. The systems were equilibrated at 300 K and evolved over 10,000 steps with a time step of 0.5~fs, totaling 5~ps of simulation time. Throughout the trajectory, the atomic structures exhibited no significant distortion or reconstruction, and the total energy per atom fluctuated stably around a well-defined mean. As shown in Figure~\ref{fig:MD-plos}, the persistence of the initial crystal frameworks after 5~ps confirms that these 2D MB$_{4}$H materials retain their structural integrity under thermal conditions.

    \subsection{Mechanical stability} 
The computed in-plane elastic constants for MB\(_{4}\)H, listed in Table~\ref{tab:elasticity}, show that \(C_{11} = C_{22}\), consistent with hexagonal lattice symmetry. The derived shear modulus \(C_{66}\) where  \(C_{66} = \frac{C_{11} - C_{12}}{2}\) and other elastic parameters satisfy the criteria for mechanical stability in two-dimensional materials—specifically, \(C_{11}C_{22} - C_{12}^2 > 0\) and \(C_{11}, C_{22}, C_{66} > 0\)—as outlined by Mouhat and Coudert~\cite{mouhat2014necessary}. These results confirm the mechanical robustness of the MB\(_{4}\)H monolayers.

   \begin{table}[tbh!]
    \centering
	\caption{In-plane elastic constants (\(C_{11}\), \(C_{22}\)) and the off-diagonal elastic constant (\(C_{12}\)) for various 2D MB\(_{4}\)H compounds, expressed in N/m.}
	   \begin{tabular}{|c|c|c|c|}
        \hline
		2D compounds & $C_{11}$, $C_{22}$ (N/m) & $C_{12}$ (N/m) & $C_{66}$ (N/m) \\
	   \hline
		BeB$_{4}$H & 279.81 & 156.80 & 61.51 \\
		MgB$_{4}$H & 247.67 & 128.18 & 59.75 \\
		CaB$_{4}$H & 211.39 & 97.76 & 56.82 \\
		AlB$_{4}$H & 286.28 & 144.42 & 70.93 \\
            \hline
	   \end{tabular}
	\label{tab:elasticity}
	\end{table}

    Furthermore, this robustness is supported by additional indicators of structural reliability: phonon dispersion spectra show no imaginary frequencies, indicating dynamical stability; \textit{ab initio} molecular dynamics simulations demonstrate thermal resilience; and energetic analyses yield negative relative formation energies. Collectively, these findings suggest that 2D MB\(_{4}\)H materials are not only stable but also promising candidates for experimental realization and potential applications.
\section{Electronics, phonons, and superconductivity}
\subsection{Electronics}\label{Electronics-section}
\begin{figure}[tbh!]
        \centering
            \includegraphics[width=16.5cm]{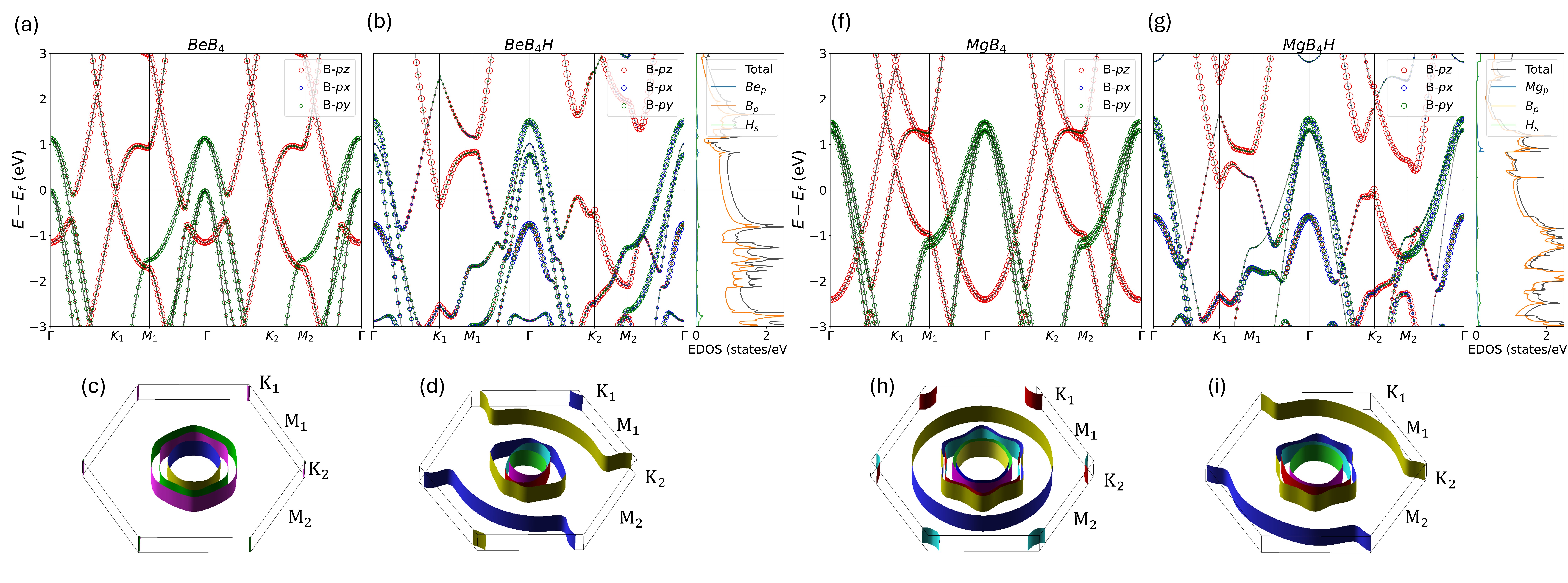}
        \caption{Figures show the electronic properties of MB$_{4}$ (left) and MB$_{4}$H compounds (right)  near the Fermi level, (M = Be, Mg) illustrating the orbital-resolved electronic band structures, the electronic density of states, the orbital-projected density of states, and the Fermi surface.}
		\label{fig:electronics-A}
    \end{figure}

\begin{figure}[tbh!]
        \centering
            \includegraphics[width=16.5cm]{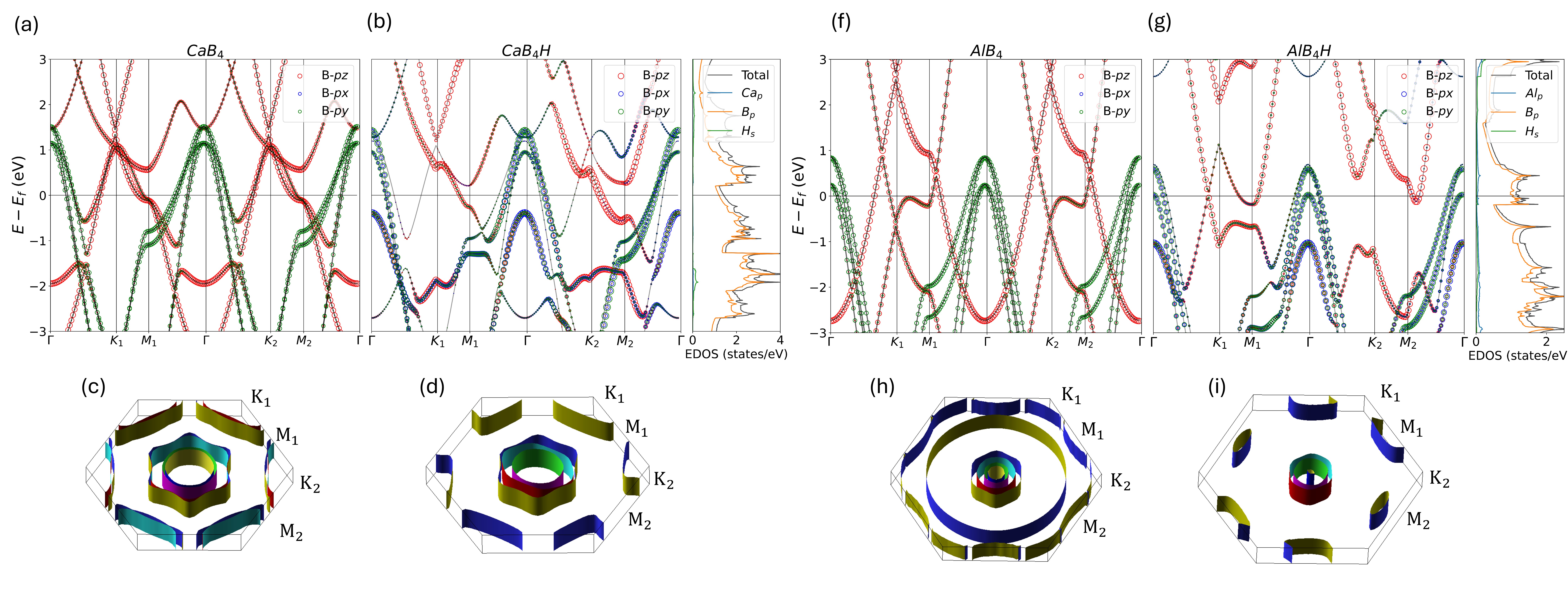}
        \caption{Figures shows electronic properties of MB$_{4}$ and MB$_{4}$H compounds (M = Ca, Al) close to the Fermi level illustrating the orbital-resolved electronic band structures, the electronic density of states, the orbital projected density of states, and the Fermi surface.}
		\label{fig:electronics-B}
    \end{figure}

The electronic properties of the pristine 2D MB$_{4}$ and MB$_{4}$H compounds (M = Be, Mg, Ca, Al) are analyzed based on DFT calculations of the Kohn-Sham electronic structure. These include the orbital-resolved electronic band structures, the electronic density of states, the orbital-projected density of states, and the Fermi surface in the Brillouin zone, as shown in Fig.~\ref{fig:electronics-A} and Fig.~\ref{fig:electronics-B}. In general, the pristine 2D trilayer MB$_{4}$ and the hydrogenated 2D MB$_{4}$H share common electronic characteristics, primarily governed by the $p$ orbitals ($p_z$, $p_x$, and $p_y$) of boron atom. These compounds exhibit metallic properties, with continuous $p$-orbital-dominated electronic bands at the Fermi level. The intersection of these bands gives rise to different Fermi surface topologies. Here, we classify the discussion into two parts: (i) Fig.~\ref{fig:electronics-A} 2D MB$_{4}$ and MB$_{4}$H compounds with M = Be, Mg, and (ii) Fig.~\ref{fig:electronics-B} 2D MB$_{4}$ and MB$_{4}$H compounds with M = Ca, Al.

The electronic band structures of BeB$_4$ and BeB$_4$H are shown in Figures~\ref{fig:electronics-A} (a) and (b), while Figures~\ref{fig:electronics-A} (f) and (g) present those of MgB$_4$ and MgB$_4$H. For BeB$_4$H and MgB$_4$H, hydrogenation alters the band dispersion, resulting in different Kohn-Sham states crossing the Fermi level. This indicates changes in charge carrier dynamics and potential modifications in electronic transport properties. As a result, the Fermi surface topologies of BeB$_4$ and BeB$_4$H, shown in Figures~\ref{fig:electronics-A} (c) and (d), along with those of MgB$_4$ and MgB$_4$H, depicted in Figures~\ref{fig:electronics-A} (h) and (i), change significantly upon hydrogenation. These modifications can directly impact superconducting properties and carrier mobility, particularly influencing anisotropic transport behaviors, as both BeB$_4$H and MgB$_4$H exhibit an one-direction open Fermi surface topology across the Brillouin zone.

    The electronic band structures of CaB$_4$ and CaB$_4$H, as shown in Figures~\ref{fig:electronics-B} (a) and (b), along with those of AlB$_4$ and AlB$_4$H in Figures~\ref{fig:electronics-B} (f) and (g), exhibit significant changes upon hydrogenation. For CaB$_4$ and CaB$_4$H, the missing of electronic bands in CaB$_4$H due to the hydrogenation results in less number of Fermi surface, but still give similar topology. For AlB$_4$ and AlB$_4$H, the Fermi surface indicate significant changes in topology upon hydrogenation, reflecting alterations in charge carrier dynamics. The emergence of two Fermi pockets and modifications in their shape suggest potential impacts on transport properties, including electrical conductivity and superconducting behavior. Interestingly, For AlB$_4$H, we observe a linear band crossing near the $K_1$ point along the $\Gamma$–$K_1$ direction, close to the Fermi level, resembling a Dirac-like feature as shown in Figure~\ref{fig:electronics-B}~(g). Although there is no Fermi surface at this particular point, this band crossing nonetheless contributes to the superconducting gap, as we will discuss further. These findings underscore the role of hydrogen in tuning the electronic structure, which could influence material properties relevant to superconductivity and electronic transport applications.

\subsection{Charge Transfer and Bonding Characteristics}
\begin{table}[htbp]
\centering
\caption{Löwdin charge analysis, pseudopotential valence, and charge transfer ($\Delta q$) in MB$_4$H monolayers. 
$\Delta$q is calculated relative to pseudopotential valence electrons. Positive $\Delta$q indicates electron gain (acceptor, $\delta^-$), negative indicates electron loss (donor, $\delta^+$).}
\begin{tabular}{|c|c|c|c|c|c|}
\hline
System & Atom & Valence (Configuration) & Total (Löwdin) & $\Delta q$ (e) & Role \\
\hline
BeB$_4$H & Be & 4.00 (1s$^2$2s$^2$) & 2.345 & $-1.655$ & Donor (Be$^{\delta+}$) \\
         & B (upper) & 3.00 (2s$^2$2p$^1$) & 3.244 & +0.244 & Acceptor (B$^{\delta-}$) \\
         & B (lower) & 3.00 (2s$^2$2p$^1$) & 3.075 & +0.075 & Slight Acceptor (B$^{\delta-}$) \\
         & H & 1.00 (1s$^1$) & 0.717 & $-0.283$ & Donor (H$^{\delta+}$) \\
\hline
MgB$_4$H & Mg & 10.00 (2p$^6$3s$^2$) & 8.380 & $-1.620$ & Donor (Mg$^{\delta+}$) \\
         & B (upper) & 3.00 (2s$^2$2p$^1$) & 3.264 & +0.264 & Acceptor (B$^{\delta-}$) \\
         & B (lower) & 3.00 (2s$^2$2p$^1$) & 2.985 & $-0.015$ & Slight Donor (B$^{\delta+}$) \\
         & H & 1.00 (1s$^1$) & 0.727 & $-0.273$ & Donor (H$^{\delta+}$) \\
\hline
CaB$_4$H & Ca & 10.00 (2p$^6$3s$^2$) & 8.363 & $-1.637$ & Donor (Ca$^{\delta+}$) \\
         & B (upper) & 3.00 (2s$^2$2p$^1$) & 3.274 & +0.274 & Acceptor (B$^{\delta-}$) \\
         & B (lower) & 3.00 (2s$^2$2p$^1$) & 2.970 & $-0.030$ & Slight Donor (B$^{\delta+}$) \\
         & H & 1.00 (1s$^1$) & 0.743 & $-0.257$ & Donor (H$^{\delta+}$) \\
\hline
AlB$_4$H & Al & 11.00 (2p$^6$3s$^2$3p$^1$) & 9.906 & $-1.094$ & Slight Donor (Al$^{\delta+}$) \\
         & B (upper) & 3.00 (2s$^2$2p$^1$) & 3.245 & +0.245 & Acceptor (B$^{\delta-}$) \\
         & B (lower) & 3.00 (2s$^2$2p$^1$) & 3.081 & +0.081 & Slight Acceptor (B$^{\delta-}$) \\
         & H & 1.00 (1s$^1$) & 0.734 & $-0.266$ & Donor (H$^{\delta+}$) \\
\hline
\end{tabular}
\label{tab:charges}
\end{table}

Löwdin charge analysis as shown in Table~\ref{tab:charges} provides insight into the charge redistribution and bonding nature of the MB$_4$H (M = Be, Mg, Ca, Al) monolayers. In all systems, a consistent charge transfer is observed from the metal atoms toward the boron and hydrogen atoms, highlighting the donor role of the metals. Be, Mg, and Ca act as strong electron donors, losing approximately 1.6--1.7 e$^-$ relative to their pseudopotential valence electrons, while Al behaves as a mild donor, losing about 1.1 e$^-$. The upper B atoms exhibit electron gain of around +0.24--0.27 e$^-$, indicating their role as acceptors, whereas the lower B atoms show only a small charge variation, reflecting their partial participation in charge redistribution. Hydrogen atoms slightly donate electrons ($\Delta q \approx -0.27$ e$^-$), further balancing the charge transfer. The total Löwdin charge does not completely recover the total valence electron count, implying that a portion of electron density remains delocalized within the M–B and B–B bonds. Orbital population analysis indicates that the lower B atoms possess an s/p ratio of approximately 0.35, consistent with sp$^2$ hybridization, suggesting trigonal planar bonding within the boron layer. This configuration supports covalent B–B bonding accompanied by delocalized $\pi$-electrons, which is characteristic of boron-based two-dimensional networks. 

     \begin{figure}[tbh!]
        \centering
        \includegraphics[width=9cm]{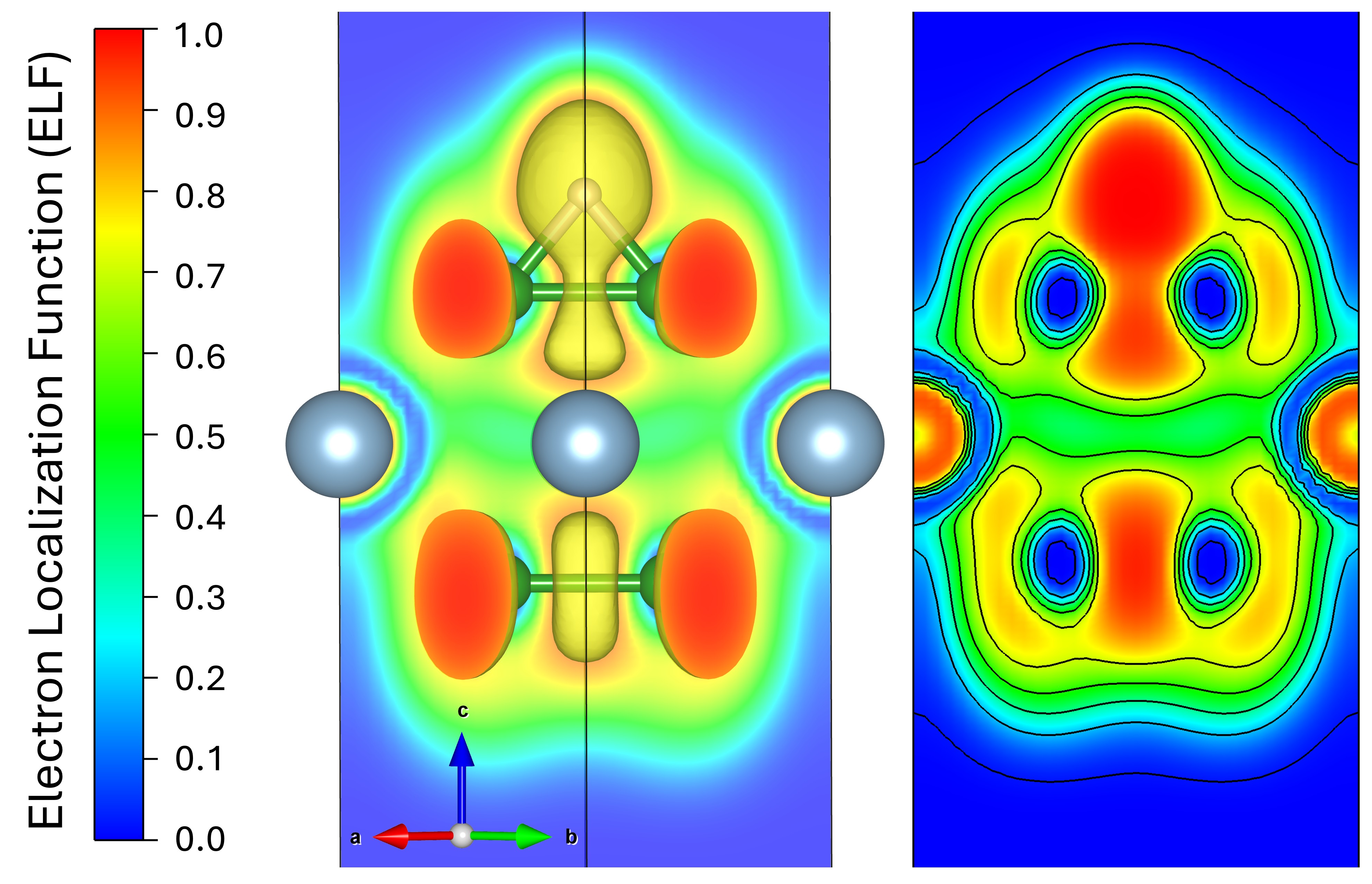}
    \caption{Electron localization function (ELF) of MB$_4$H monolayer. High ELF values (red, ELF $>$ 0.8) between the lower B atoms indicate strong sp$^2$ covalent B--B bonds with delocalized $\pi$ electrons. Intermediate ELF values (0.5--0.7) between the upper B and metal atoms reveal mixed ionic--covalent M--B interactions due to partial electron donation from the metal to boron. The absence of an ELF attractor between B and H atoms suggests predominantly ionic B--H bonding, where H acts as an electron donor.}
        \label{fig:ELF_MB4H}
    \end{figure}

The bonding characteristics are further confirmed by the electron localization function (ELF), as shown in Fig.~\ref{fig:ELF_MB4H}. The ELF shows strong localization (ELF $>$ 0.8) between the lower B atoms, confirming  covalent bonding within the boron plane. In contrast, the absence of a distinct ELF attractor between boron and hydrogen atoms suggests an ionic B–H interaction, with hydrogen acting as an electron donor. Regions surrounding the metal atoms show low ELF values (ELF $\approx$ 0.3--0.5), consistent with partial ionic character. Within the boron layer,  intermediate ELF values (0.5--0.7) indicate the presence of delocalized $\pi$ electrons contributing to covalency and metallic behavior. 

The ELF map also differentiates the boron sublayers. A strong ELF attractor (ELF $>$ 0.8)  between the B atoms in the lower layer corresponds to sp$^2$ covalent bonding with delocalized $\pi$ electrons. In contrast, the upper B atoms exhibit intermediate ELF values (0.5--0.7) in the region between the metal and boron, with the localization basin is polarized toward boron. 
Together the Löwdin charge and ELF analyses consistently reveal a mixed ionic--covalent bonding character: the lower boron atoms form sp$^2$ covalent bonds with delocalized $\pi$ electrons within the boron plane, the upper boron atoms show polar ionic--covalent interactions with the metal atoms, and the B--H bond exhibit predominantly polar covalent character.

\subsection{Phonon-mediated superconductivity}
     \begin{figure}[tbh!]
        \centering
        \includegraphics[width=17cm]{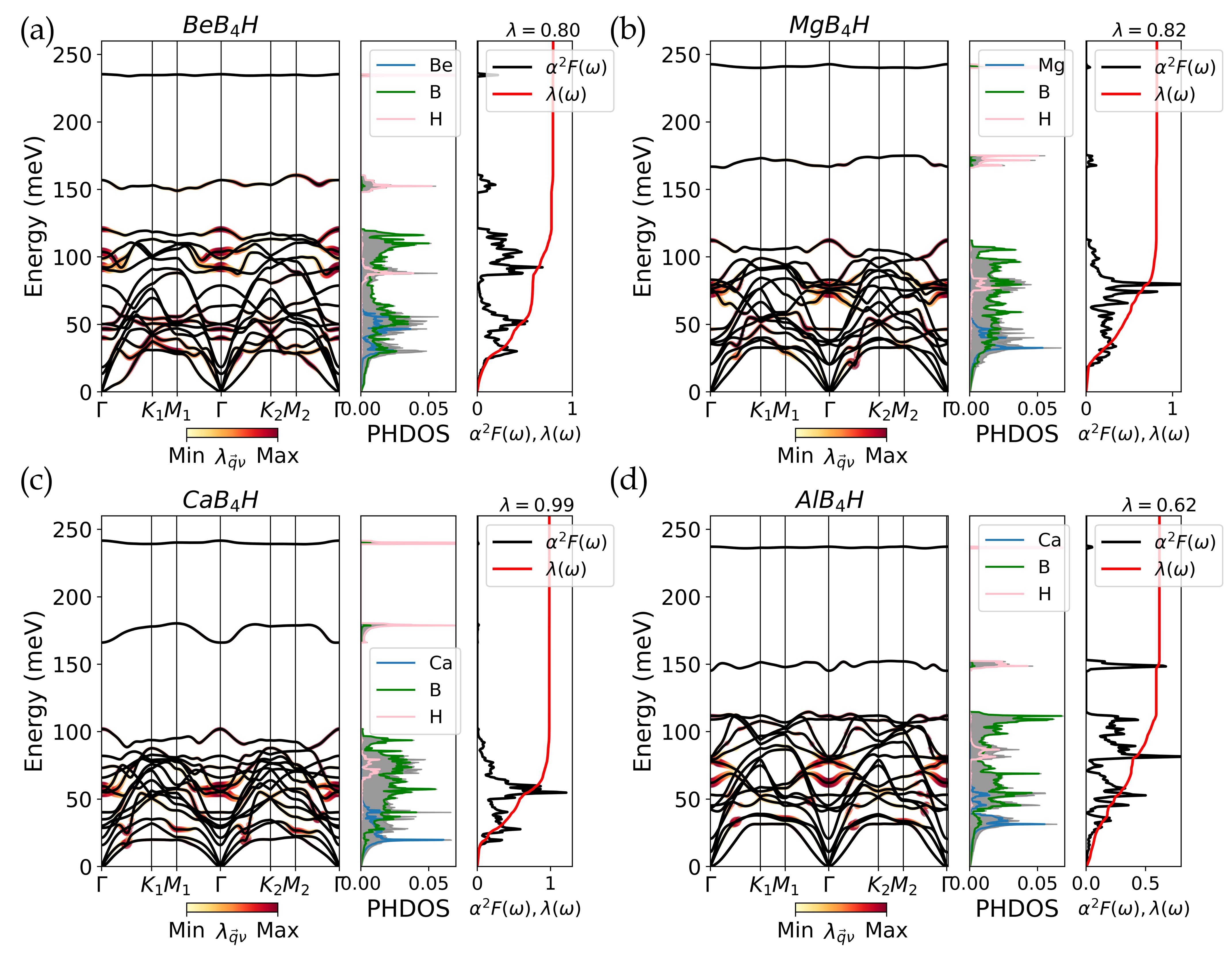}
        \caption{(a-d) show the phonon properties and electron-phonon interaction of MB$_{4}$H compounds (M = Be, Mg, Ca, Al). It consists of the phonon dispersion, phonon density of states, the isotropic Eliashberg spectral function $\alpha^2 F (\omega)$ and $\lambda (\omega)$.}
		\label{fig:phband-phdos-a2f}
    \end{figure}
    The phonon dispersion relations, projected phonon density of states (PHDOS), and Eliashberg spectral functions $\alpha^2F(\omega)$ for BeB$_4$H, MgB$_4$H, CaB$_4$H, and AlB$_4$H are presented in Figure~\ref{fig:phband-phdos-a2f}. The phonon dispersion curves in Figures~\ref{fig:phband-phdos-a2f}(a)--(d) illustrate the vibrational modes along high-symmetry directions in the Brillouin zone. The absence of imaginary frequencies in all four compounds confirms their dynamical stability. The PHDOS plots reveal that hydrogen atoms predominantly contribute to the high-frequency phonon modes around 150~meV and 250~meV, while the low-frequency modes (0--100~meV) are mainly associated with metal and boron vibrations.

    The Eliashberg spectral function $\alpha^2F(\omega)$ and the frequency-dependent electron-phonon coupling constant $\lambda(\omega)$, shown in the rightmost panels of Figure~\ref{fig:phband-phdos-a2f}, provide further insights into the superconducting behavior. The calculated total electron-phonon coupling constants $\lambda$ are 0.80, 0.82, 0.99, and 0.62 for BeB$_4$H, MgB$_4$H, CaB$_4$H, and AlB$_4$H, respectively. The relatively high $\lambda$ values in BeB$_4$H, MgB$_4$H, and CaB$_4$H suggest strong electron-phonon interactions conducive to superconductivity. In contrast, the lower $\lambda$ value in AlB$_4$H indicates weaker coupling. The spectral function $\alpha^2F(\omega)$ reveals that the dominant phonon contributions to electron-phonon interaction arise in the low-to-mid frequency range, primarily involving metal and boron vibrations. The sharp increase in $\lambda(\omega)$ at low frequencies further supports the conclusion that electron-phonon coupling is largely governed by metal–boron vibrational modes.

The mode-resolved electron--phonon coupling $ \lambda_{\vec{q}\nu} $, shown in Figure~5, highlights the momentum- and mode-dependent contributions to the total electron--phonon coupling constant $ \lambda $, which is derived from the Eliashberg spectral function 
$ \alpha^2F(\omega) $.

For $\mathrm{BeB}_4\mathrm{H}$, two frequency ranges  dominate  $ \lambda $: 30-60meV  and 90-110meV.  
In the 30-60meV range, the largest values of $ \lambda_{\vec{q}\nu} $ occur near  to the $\mathrm{K}_1$ and $\mathrm{M}_1$ points, with smaller contributions from $\mathrm{K}_2$ and $\mathrm{M}_2$. This behavior arises from interband Fermi surface nesting along the $\mathrm{K}_1$ and $\mathrm{M}_1$ directions between the outer asymmetric Fermi surface sheets (Figure~3b).
A similar effect is observed in $\mathrm{MgB}_4\mathrm{H}$, where the low phonon energy leads to a low superconducting gap energy. 

In contrast, for CaB$_4$H and AlB$_4$H the asymmetry  along the $\mathrm{K}_1 \rightarrow \mathrm{M}_1 \rightarrow \Gamma \rightarrow \mathrm{K}_2 \rightarrow \mathrm{M}_2$ path is less pronounced. This difference arises from the presence of Fermi surface sheets along these high-symmetry directions.

The existence of a Fermi surface away from the $\Gamma$ point along the $\mathrm{K}_1 \rightarrow \mathrm{M}_1 \rightarrow \Gamma \rightarrow \mathrm{K}_2 \rightarrow \mathrm{M}_2$ path leads to finite values of $ \lambda_{\vec{q}\nu} $ in regions distant from $\Gamma$. Consequently, Cooper pairs can form on Fermi surface regions far from the $\Gamma$ point, resulting in the low-energy superconducting gap observed in Figure~6 for these compounds.
Finally, a small phonon softening is observed as a dip in the phonon dispersion of MgB$_4$H, which becomes even more pronounced in CaB$_4$H.

In the higher-energy range of 90--110~meV, $\lambda_{\vec{q}\nu}$ exhibits strong peaks near the $\Gamma$ point, primarily associated with phonon modes 13 and 15. These contributions result mainly from intraband ($k \rightarrow k$) transitions within the same electronic band, as the phonon wavevector approaches zero. Similar behavior is observed across other metal hydrides: in MgB$_4$H, phonon modes 10, 13, and 14 display strong coupling at $\Gamma$, while in CaB$_4$H, modes 10, 11, and 12 contribute significantly. These zone-center optical phonons strongly couple to electronic states at the Fermi level, leading to a high superconducting gap for electrons at the Fermi surface near these electron momentum states in all four cases. However, for MgB$_4$H and CaB$_4$H, the superconducting gap energy is weak on the inner Fermi surface, which could potentially result from a small phonon softening which is observed as a dip in the phonon dispersion of MgB$_4$H and CaB$_4$H.

    \subsection{Multigap Superconductivity}
    \begin{figure}[tbh!]
        \centering
        \includegraphics[width=10cm]{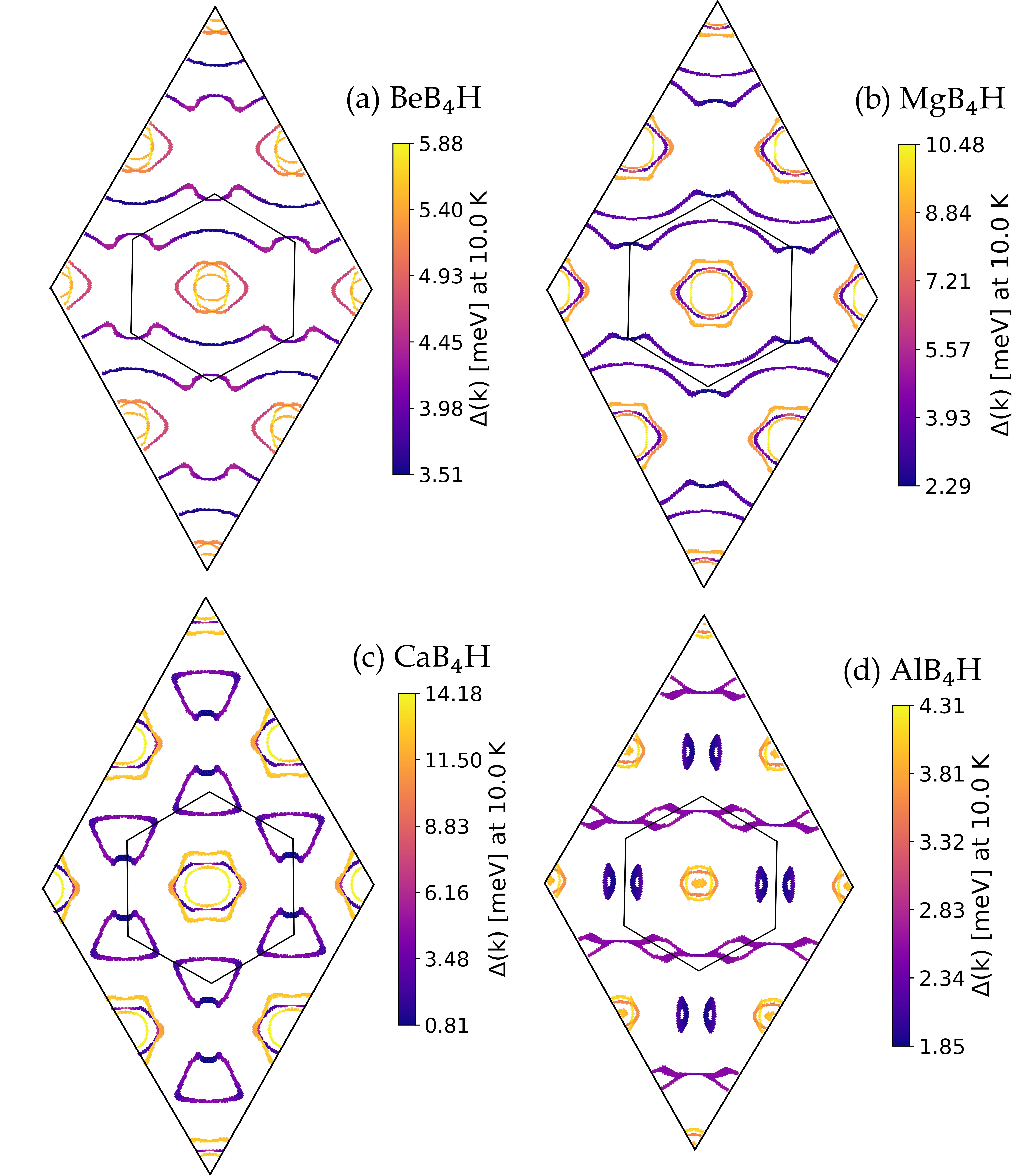}
        \caption{(a-d) show contour anisotropic superconducting gap ($\Delta_{nk}(T)$) of MB$_{4}$H compounds (M = Be, Mg, Ca, Al) in Brillouin zone at T=10~K.}
		\label{fig:contour-gap}
    \end{figure}

\begin{figure}[tbh!]
        \centering
        \includegraphics[width=10cm]{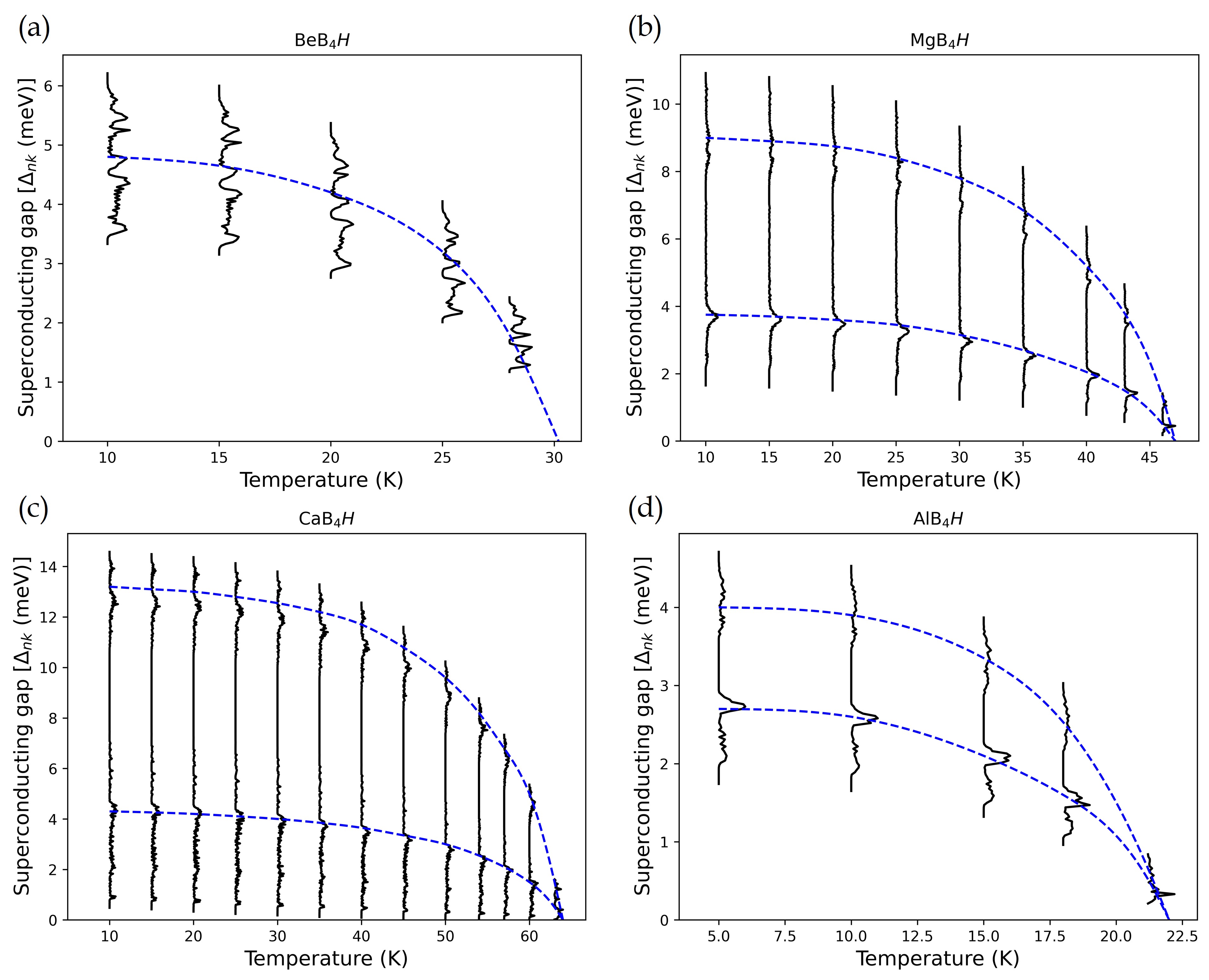}
        \caption{(a-d) show anisotropic superconducting gap ($\Delta_{nk}(T)$) of MB$_{4}$H compounds (M = Be, Mg, Ca, Al) in the unit of meV. The gaps graudally are decreasing with increasing temperature and vanish at certain critical temperatures.}
		\label{fig:sc_gap}
    \end{figure}

    The temperature-dependent anisotropic superconducting gaps $\Delta_{nk}(T)$ for BeB$_4$H, MgB$_4$H, CaB$_4$H, and AlB$_4$H are presented in Figure~\ref{fig:sc_gap}. The superconducting gaps exhibit a clear temperature dependence, gradually decreasing as the temperature increases and vanishing at the critical temperature ($T_c$), following the conventional BCS-like superconducting behavior. However, the evolution of the superconducting gaps also reveals significant anisotropy, as evidenced by the fluctuating black curves in Figures~\ref{fig:sc_gap} (a)-(d), which show distinct gap values for different electronic states.  

    The anisotropic nature of the superconducting gap indicates that these materials exhibit multiple superconducting gap values across different momentum states, a defining characteristic of two-gap superconductivity. This behavior is particularly evident in MgB$_4$H, CaB$_4$H, and AlB$_4$H, where the variation in $\Delta_{nk}$ values includes regions with vanishing $\Delta_{nk}$ in the energy spectrum at each temperature. For BeB$_4$H, no multi-gap superconducting behavior is observed due to the smooth variation of the superconducting gap, as shown in the contour plot in Figure~\ref{fig:contour-gap}(a). For the other materials, two distinct regions are observed in the contour plots of the superconducting gap, as shown in Figures~\ref{fig:contour-gap}~(b, c, d). These findings are consistent with previous investigations of their pristine counterparts: CaB$_4$ has been predicted to exhibit three-gap superconductivity~\cite{sevik2022high}, MgB$_4$ four-gap superconductivity~\cite{zhao2020mgb}, and AlB$_4$ three-gap superconductivity~\cite{zhao2019two}. However, due to the distortion introduced by hydrogenation, the three-, four-, and three-gap superconductivity of CaB$_4$, MgB$_4$, and AlB$_4$ is reduced to two-gap superconductivity, accompanied by an enhancement in the critical temperature, as summarized in Table~\ref{tab:superconducting}. As shown in Figure~\ref{fig:contour-gap}~(d), a superconducting gap emerges near the $K_1$ point (at the top and bottom of the hexagon). Although there is no Fermi surface at this location, as shown in Figure~\ref{fig:electronics-B}, these superconducting gaps originate from a linear band crossing near the $K_1$ point along the $\Gamma$–$K_1$ direction, close to the Fermi level, resembling a Dirac-like feature, as illustrated in Figure~\ref{fig:electronics-B}~(g). Therefore, similar to other known two-gap superconductors—such as n-doped graphene \cite{margine2014two}, AlB$_2$-based thin films \cite{zhao2019two}, trilayer LiB$_2$C$_2$ \cite{gao2020strong}, monolayer LiBC \cite{modak2021prediction}, GaInSLi~\cite{seeyang2025phase_japtwpgap}, MoSLi \cite{xie2024strong}, MoSeLi \cite{seeyangnok2024twomoseli}, and MoSH \cite{liu2022two}—the MgB$_4$H, CaB$_4$H, and AlB$_4$H systems also exhibit intrinsic multigap superconductivity. 
    
    Among the studied compounds, CaB$_4$H exhibits the largest superconducting gap values, reaching approximately 14 meV at low temperatures, whereas AlB$_4$H shows the smallest gap, in agreement with its lower electron-phonon coupling constant. This correlation between the superconducting gap magnitude and electron-phonon interaction strength further supports the role of phonon-mediated pairing in determining $T_c$ and gap anisotropy. With the largest superconducting gap and electron-phonon coupling constant, CaB$_4$H has an intrinsic superconducting temperature ($T_c$) of 64 K. The second-largest superconducting gap, approximately 10 meV, corresponds to CaB$_4$H with $T_c = 47$ K. The two lowest superconducting gaps belong to BeB$_4$H and AlB$_4$H, with intrinsic superconducting temperatures of $T_c = 30$ K and $T_c = 22$ K, respectively. Table~\ref{tab:superconducting} summarize electron-phonon coupling constant ($\lambda$) and superconducting temperature (T$_c$) of the pristine 2D MB$_{4}$ and MB$_{4}$H compounds (M = Be, Mg, Ca, Al).

    \begin{table}[tbh!]
    \centering
	\caption{The table shows the comparison of electron-phonon coupling constant ($\lambda$) and superconducting temperature (T$_c$) between the pristine 2D MB$_{4}$ and MB$_{4}$H compounds (M = Be, Mg, Ca, Al).}
	   \begin{tabular}{|c|c|c|c|c|c|c|c|c|c|}
        \hline
		Pristine & $\lambda$ & T$_c^{ME}$ (K) & Hydrogenation & $\lambda$ & $\omega_{\text{log}}$(meV) & $\omega_{2}$(meV) & T$_c^{ME}$ (K) \\
	   \hline
		BeB$_{4}$ & 1.212 & 29.9 \cite{sevik2022high} & BeB$_{4}$H & 0.80 & 43.94 & 64.94 & 30 \\
		MgB$_{4}$ & 0.808 & 22.2 \cite{sevik2022high} & MgB$_{4}$H & 0.82 & 45.38 & 58.39 & 47 \\
		CaB$_{4}$ & 1.196 & 36.1 \cite{sevik2022high} & CaB$_{4}$H & 0.99 & 38.04 &  47.10 & 64 \\
		AlB$_{4}$ & 0.911 & 30.9 \cite{sevik2022high} & AlB$_{4}$H & 0.62 & 48.30 & 71.50 & 22 \\
            \hline
	   \end{tabular}
	\label{tab:superconducting}
	\end{table}

    Overall, these results highlight the interplay between temperature dependence, gap anisotropy, and multi-gap superconducting behavior in MB$_4$H compounds. The tunability of the superconducting gap structure via metal substitution suggests that these materials may serve as promising candidates for engineered superconducting applications, where controlling anisotropy and multi-gap features can lead to optimized superconducting properties.  

\section{Conclusion} \label{Sec:Conclusion}
In this study, we have systematically explored the electronic structure, phonon properties, and superconducting behavior of two-dimensional hydrogenated trilayer metal borides (MB$_4$H; M = Be, Mg, Ca, Al). Our results reveal that these compounds share common electronic characteristics, including superconductivity predominantly governed by the $p$ orbitals ($p_z$, $p_x$, and $p_y$) of boron. The metallic nature of these materials arises from continuous $p$-orbital-dominated electronic bands at the Fermi level, leading to distinct Fermi surface topologies. While hydrogenation significantly modifies the band dispersion in BeB$_4$H and MgB$_4$H, resulting in an open Fermi surface topology, AlB$_4$H exhibits more complex changes with electron pockets, while CaB$_4$H retains a similar topology to pristine CaB$_4$ with a smaller number of Fermi surfaces. The emergence of Fermi pockets in AlB$_4$H suggests possible implications for its transport and superconducting properties.


Phonon calculations confirm the dynamical stability of these materials and highlight strong electron-phonon interactions, which play a key role in their superconducting properties. The temperature-dependent evolution of the anisotropic superconducting gap indicates multi-gap superconductivity, likely influenced by the intricate Fermi surface structure. Among the studied compounds, CaB$_4$H exhibits the highest superconducting gap and strongest electron-phonon coupling, yielding an intrinsic superconducting temperature ($T_c$) of 64 K. In contrast, AlB$_4$H shows the weakest coupling, with $T_c = 22$ K, underscoring the tunability of superconducting properties through elemental substitution. The calculated electron-phonon coupling constants ($\lambda$) are 0.80, 0.82, 0.99, and 0.62 for BeB$_4$H, MgB$_4$H, CaB$_4$H, and AlB$_4$H, respectively.

The effect of hydrogenation on T$_c$ is dramatic, but varied.  In MgB$_4$ and CaB$_4$ the $T_c$ is doubled, in BeB$_4$, it barely changes, while in AlB$_4$ it is reduced.

Overall, our findings offer a comprehensive understanding of the interplay between electronic, phononic, and superconducting properties in MB$_4$H compounds. The strong electron-phonon coupling, anisotropic superconducting gap structure, and multi-gap nature of these materials suggest their potential for superconducting applications. The large effect of hydrogenation on T$_c$ suggests that doping is an extremely effective way to tune the $T_c$ of 2D superconductors

    \section*{Data Availability}
    The data that support the findings of this study are available from the corresponding
    authors upon reasonable request.
    
    \section*{Code Availability}
    The first-principles DFT calculations were performed using the open-source Quantum ESPRESSO package, available at \url{https://www.quantum-espresso.org}, along with pseudopotentials from the Quantum ESPRESSO pseudopotential library at \url{https://pseudopotentials.quantum-espresso.org/}. Electron-phonon coupling and related properties were computed using the EPW code, available at \url{https://epw-code.org/}.

    \section*{Acknowledgements}
	This research project is supported by the Second Century Fund (C2F), Chulalongkorn University. We acknowledge the supporting computing infrastructure provided by NSTDA, CU, CUAASC, NSRF via PMUB [B05F650021, B37G660013] (Thailand). (\url{URL:www.e-science.in.th}). This also work used the ARCHER2 UK National Supercomputing Service (\url{https://www.archer2.ac.uk}) as part of the UKCP collaboration.

    \section*{Author Contributions}
    Jakkapat Seeyangnok performed all of the calculations, analysed the results, wrote the first draft manuscript, and coordinated the project. Udomsilp Pinsook analysed the results and wrote the manuscript. Graeme Ackland analysed the results, and wrote the final manuscript. All authors have approved the final manuscript.

    \section*{Conflict  of Interests}
    The authors declare no competing financial or non-financial interests.

    \bibliographystyle{unsrt} 
	\bibliography{references}

\end{document}